\documentclass[11pt,a4paper]{article}
\usepackage{hyperref}
\usepackage{amsmath,amssymb}
\usepackage{dsfont}
\usepackage{graphicx}
\usepackage[numbers,sort]{natbib}
\usepackage[boxed]{algorithm2e}

\usepackage{anysize}
\marginsize{2cm}{2cm}{2cm}{2cm}

\newcommand{\vect}[1]{\mathbf{\boldsymbol{#1}}}

\newcommand{\order}[0]{\sqsubset}
\newcommand{\orders}[1]{\sqsubset_{#1}}
\newcommand{\DP}[0]{\textrm{DP}}
\newcommand{\GEM}[0]{\textrm{GEM}}
\newcommand{\pa}[0]{\textrm{pa}}
\newcommand{\araba}[0]{\textit{A. thaliana }}
\newcommand{\droso}[0]{\textit{D. melanogaster }}

\begin{document}
\title{Inference of Temporally Varying Bayesian Networks}
\author{Thomas Thorne \& Michael P.H. Stumpf\\
Centre of Integrative Systems Biology and Bioinformatics,\\ Division of Molecular Biosciences, Imperial College London,\\ London SW7 2AZ, UK}
\date{\today}
\maketitle

\begin{abstract}
When analysing gene expression time series data an often overlooked but crucial aspect of the model is that the regulatory network structure may change over time. Whilst some approaches have addressed this problem previously in the literature, many are not well suited to the sequential nature of the data. Here we present a method that allows us to infer regulatory network structures that may vary between time points, utilising a set of hidden states that describe the network structure at a given time point. To model the distribution of the hidden states we have applied the Hierarchical Dirichlet Process Hidden Markov Model, a nonparametric extension of the traditional Hidden Markov Model, that does not require us to fix the number of hidden states in advance. We apply our method to existing microarray expression data as well as demonstrating is efficacy on simulated test data.
\end{abstract}

\section{Introduction}

The analysis of gene expression data in the field of systems biology is a challenging problem for a number of reasons, not least because of the high dimensionality of the data and relative dearth of data points. A number of approaches have been taken to inferring regulatory interactions from such data, often employing graphical models or sparse regression techniques \citep{Schafer:2005kl,OpgenRhein:2007ih,Lebre:2009ea}. 

These problems are further compounded by the nature of the biological systems under consideration, due to the influence of unobserved actors that may alter the behaviour of the system. Often experiments are performed over long time periods during which it is natural to expect the regulatory interactions at work to change. It is also worth noting that the time scales of regulatory responses to stimuli often differ from those of signalling and metabolic responses, and so it may be that responses to stimuli, around which experiments are often designed, take place in several phases each having different time scales.

Previous studies have attempted to address this problem by introducing changepoints in the time series, allowing the inferred network structure to differ between the resulting segments of the time series. For example in \citet{Lebre:2010jz} a changepoint model is applied in which Dynamic Bayesian Networks are inferred for each segment of the time series. However such approaches may place strong prior assumptions on the number of changepoints that may be observed, and do not adjust for the complexity of the observed data automatically. Instead in \citet{Grzegorczyk:2008ie} an allocation sampler is used in combination with Bayesian Networks to assign each observation to a group, but unlike changepoint models this method treats the observations as being exchangeable, ignoring the fact that the data are sequential. The similar methodology in \citet{Ickstadt2011} uses a more flexible nonparametric prior on group assignments, applied to the modelling of molecular interactions using Bayesian Networks, but suffers the same drawbacks in not recognising the sequential nature of the data.

Here we present a methodology that allows us to infer network structures that may change between observations in a nonparametric framework whilst modelling the sequential nature of the data. To that end we have employed the infinite hidden Markov model (iHMM) of \citet{Beal:2002tt}, also known as the hierarchical Dirichlet Process Hidden Markov Model (HDP-HMM) \citep{Teh:2010to}, in particular the ``Sticky'' extension of \citet{Fox:2009kf}, in conjunction with a Bayesian network model of the gene regulatory network structure. The HDP-HMM allows the number of different states of the network structure to adapt as necessary to explain the observed data, including a potentially infinite number of states, of course restricted in practice by the finite number of experimental observations. In the previous work of \citet{2010arXiv1001.4208R} it was demonstrated that the HDP-HMM outperforms a Dirichlet Process mixture for Gaussian graphical models on heterogeneous time series.

We apply our methodology to both simulated data and gene expression data for \textit{Arabadopsis Thaliana} and \textit{Drosophila Melanogaster}, demonstrating its effectiveness in detecting changes in network structure from time series data.

\section{Methods}

Given gene expression time series data over $m$ genes at $n$ time points, we denote the observations as the $n\times m$ matrix $\vect{X}=(\vect{x_1},\dots, \vect{x_n})^T$, where $\vect{x_j}=(x_{j1},\dots,x_{jm})^T$, the column vector of expression levels for each of the $m$ genes at time point $j$. We formulate our model as a Hierarchical Dirichlet Process Hidden Markov Model, a stochastic process whereby a set of hidden states $s_1,\dots,s_n$ govern the parameters of some emission distribution $F$ over a sequence of time points $1\dots n$.

Each observation $\vect{x_j}$ is then generated from a corresponding emission distribution $F(\theta_k)$, where $s_j=k$. For our emission distributions, $F$, we use a Bayesian Network model over the $m$ variables to represent the regulatory network structures corresponding to each hidden state.

\subsection{Hierarchical Dirichlet Process Hidden Markov Models}

To model a hidden state sequence that evolves over time we apply the methodology first introduced in \citet{Beal:2002tt} whereby a finite state Hidden Markov Model, consisting of a set of hidden states $s_1,\dots, s_n$ over some alphabet $1\dots K$, is extended so that $K \rightarrow \infty$. In a classical Hidden Markov Model, the number of states $K$ is typically specified in advance, and states follow a Markovian distribution whereby transitions are made between states with probability $\pi_{kl} = p(s_j=l|s_{j-1}=k)$, so that the next state in the sequence is drawn conditional only on the previous state.

The Hierarchical Dirichlet Process Hidden Markov Model (HDP-HMM) \citep{Beal:2002tt,Teh:2006wl,Teh:2010to} instead applies a Dirichlet Process prior to the transition probabilities $\vect{\pi}_{k\cdot}$ out of each of the states, and uses a hierarchical structure to couple the distributions between the individual states to ensure a shared set of potential states into which transitions can be made across all of the $\vect{\pi}$. This allows for an unlimited number of potential states, of course limited in practice by the number of observed data points.

More formally, each hidden state $k$ possesses a Dirichlet Process $G_k$ from which the next state is drawn, and a common (discrete) base measure $G_0$ is shared between the $G_k$, so that $G_k \sim \DP(\alpha,G_0)$. As a result transitions are made into a discrete set of states shared across all of the $G_k$, and drawn from $G_0$. The base measure $G_0$ is in turn drawn from a Dirichlet Process, $G_0 \sim \DP(\gamma,H)$, $H$ being our prior over parameters for the emission distributions $F_k$. 

\begin{figure}
\begin{center}
\includegraphics{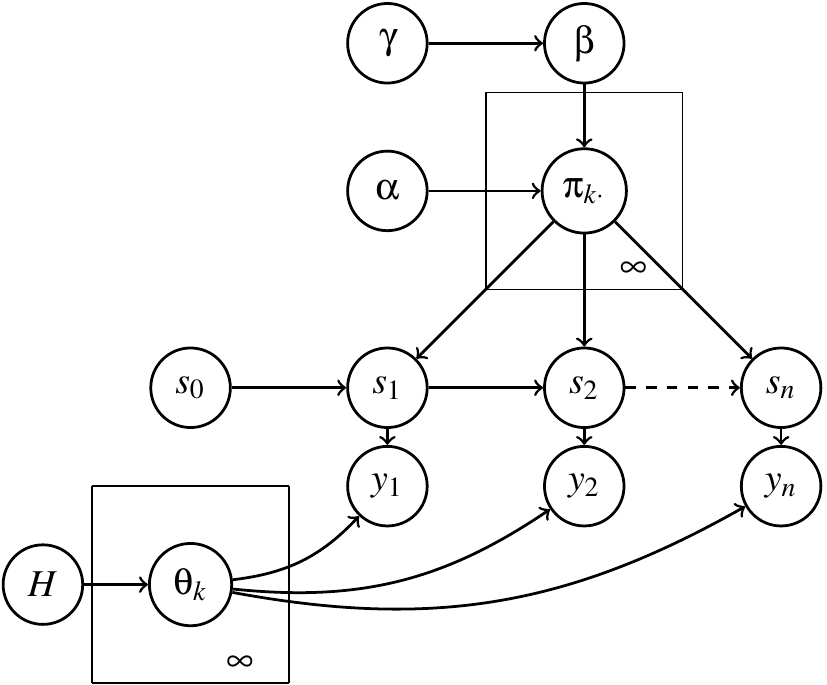}
\caption{\label{hdphmmgm} The Hierarchical Dirichlet Process Hidden Markov Model (HDP-HMM) represented as a graphical model.}
\end{center}
\end{figure}

Then using the stick breaking construction of \citet{Sethuraman:1994} for $G_0$ and drawing $\theta_l$ independently from $H$, we have that $G_0 = \sum_l^\infty \beta_{l}\delta_{\theta_l}$, with $\vect{\beta} \sim GEM(\gamma)$, and so $G_k = \sum_l^\infty \pi_{kl}\delta_{\theta_l}$ with $\vect{\pi}_k \sim \DP(\alpha,\vect{\beta})$. The resulting model is outlined in figure \ref{hdphmmgm}.

However in a biological system it is probably realistic to assume that only a subset of the wide variety of potential behaviours of the hidden state sequence is relevant, as behaviour such as rapid cycling between states at adjacent time points would \textit{a priori} seem to be unlikely to be observed in most gene expression data sets.

Thus we choose to apply the Sticky HDP-HMM \citep{Fox:2008dm,Fox:2009kf}, that introduces an extra parameter $\kappa$ that biases the prior probability of transitions between states towards remaining in the current state rather than transitioning to a differing one. Adding such a prior assumption simply states that we expect the state of the system to remain the same between successive time points; this is both parsimonious and  would seem to be justified in the case of gene expression microarray data sets, where we might only expect to observe a small number of transitions to differing states across the time series.

This modification to the HDP-HMM gives us a model generating observed data points $x_j$ as \citep{Fox:2009kf}

\begin{align}
\vect{\beta}|\gamma & \sim \GEM(\gamma),\\
\vect{\pi_{k\cdot}}| \alpha,\vect{\beta},\kappa & \sim \DP\left(\alpha+\kappa,\frac{\alpha\vect{\beta}+\kappa\delta_k}{\alpha+\kappa}\right),\\
s_j|s_{j-1},\vect{\pi} & \sim \vect{\pi}_{s_{j-1}\cdot},\\
\theta & \sim H,\\
x_j|s_j & \sim F(\theta_{s_j}).
\end{align}

\subsection{Gibbs sampling for the Sticky HDP-HMM}

To sample from the hidden state sequence we have used a Gibbs sampling procedure based on the conditional probabilities for the hidden state $s_i$ given the remaining hidden states $s_{-i}$ \citep{Fox:2008dm}, updating each hidden state individually in a sweep over the $n$ states,

\begin{equation}
\begin{split}
p(s_j=k|\vect{s}_{-j},\alpha,\beta,\kappa) \propto & (N^{-j}_{s_{j-1}k}+\alpha\beta_k+\kappa\delta_{s_{j-1}}(k))\\
& \left(\frac{N^{-j}_{ks_{j+1}}+\alpha\beta_{s_{j+1}}+\kappa\delta_{s_{j+1}}(k)+\delta_{s_{j-1}}(k)\delta_{s_{j+1}}(k)}{\alpha + N^{-j}_{k\cdot}+\kappa+\delta_{s_{j-1}}(k)}\right)\\
&p(\vect{X}_{j\cdot}|\vect{X}_{i\cdot} : s_i = k, i\neq j,F_k),
\end{split}
\end{equation}

where $\vect{s}_{-j}$ denotes the state sequence $s_1,\dots,s_n$ excluding $s_j$, $N^{-j}_{kl}$ indicates the number of observed transitions from state $k$ to state $l$ within the hidden state sequence $\vect{s}_{-j}$, and $N^{-j}_{k\cdot}$ the total number of transitions from state $k$ within $\vect{s}_{-j}$.

\subsection{Bayesian Network emission distributions}

To model the regulatory network structure corresponding to the hidden states of the HDP-HMM, we have applied a Bayesian Network methodology to capture the relationships between the genes represented in our data. Thus each hidden state has a unique Bayesian Network describing the interactions occurring between the genes at the time points corresponding to a particular state.

Bayesian Networks are probabilistic models, whereby a directed graph defines the conditional independence relationships between a set of random variables \citep{Koski:1315082}. For the model to remain consistent, the graph structure $\mathcal{G}$, with nodes $u \in N_{\mathcal{G}}$ representing random variables and directed edges $(v,u)\in E_\mathcal{G}$ representing conditional probability relationships between them, must be acyclic.

For a given Bayesian network structure $\mathcal{G}$ and model parameters $\vect{\theta}$, the joint distribution $p(\vect{X}|\mathcal{G},\vect{\theta})$ factorises as a product of local distributions for each node,

\begin{equation}
p(\vect{X}|\mathcal{G},\vect{\theta}) = \prod_{u\in N} p(\vect{x}_{\cdot u}|\pa_{\mathcal{G}}(u),\theta_u),
\end{equation}

 where for each observation the value $x_{iu}$ of a node $u$ is dependent on the values of its set of parent nodes $\pa_{\mathcal{G}}(u) = \lbrace v \in N | (v,u) \in \mathcal{G} \rbrace$ and some parameters $\theta_u$. Here we have used a BGe model \citep{Geiger:2002uj} that allows the variables to take continuous values and defines the local distributions for each observation $i\in 1,\dots, m$ of a gene $u$ as

\begin{equation}
x_{iu} \sim \mathcal{N}\left(\mu_u+\sum_{v \in \pa_{\mathcal{G}}(u)} b_{uv}(x_{iv}-\mu_v),\sigma^2_u\right),
\end{equation}

with parameters $\theta_u = \mu_u,\vect{b_u},\sigma^2_u$. Then using a conjugate Normal-Wishart prior and making certain assumptions we can calculate the local marginal likelihoods $p(\vect{x}_u|\pa_{\mathcal{G}}(u))$ as described in \citet{Geiger:2002uj} and from these derive the joint probability $p(\vect{X}|\mathcal{G})$.

Unfortunately, due to the restriction of the network structure to that of a directed acyclic graph (DAG) it is difficult to explore the space of possible network structures. Several MCMC schemes have been proposed, including those of \citet{Grzegorczyk:2008tw} and \citet{Madigan:1995vv}, but performing random walks over DAG network structures faces the problem that proposing moves that maintain the DAG structure can be complex and time consuming, and mixing of the Markov chain can be slow.

However, as noted in \citet{Friedman:2003kw}, a DAG structure $\mathcal{G}$ corresponds to a partial ordering on the nodes and so induces a (non-unique) total ordering, and so the method of \citet{Friedman:2003kw} instead proposes to perform a random walk over total orderings of the nodes. This allows the Markov chain to move quickly around the space of possible graph structures, improving the mixing properties of the chain.

Whilst this introduces a bias in the prior distribution over graph structures \citep{Grzegorczyk:2008tw} it greatly simplifies the computational complexity of the MCMC procedure and such a bias may be justified by arguments of parsimony, since graphs consistent with more orderings are more likely to be sampled. Furthermore the uniform prior on DAG structures is not uniform over Markov equivalent graphs, and so also introduces a different kind of bias in the results. Finally, a trivial modification of the algorithm of \citet{Friedman:2003kw} allows for a correction of the bias \citep{Ellis:2008ka}.

Thus in our methodology we apply the MCMC sampler of \citet{Friedman:2003kw} to infer Bayesian Network structures for each hidden state of the HDP-HMM by sampling over total orderings of the nodes $\order$ given the data points corresponding to the state in question. It is easy to calculate the likelihood of an ordering $\order$ using \citep{Friedman:2003kw} 

\begin{equation}
p(\vect{X}|\order) = \prod_{u\in N_{\mathcal{G}}} \sum_{k\in\pa_{\mathcal{G}}^{\order}(u)} p(\vect{x}_{\cdot u},k),
\end{equation}

where $\pa_{\mathcal{G}}^{\order}$ denotes the set of possible parent sets over the nodes of $\mathcal{G}$ consistent with the ordering $\order$. Then we can use a Metropolis Hastings sampler to sample from the posterior of orderings $p(\order|\vect{X})=p(\vect{X}|\order)p(\order)$ \citep{Ellis:2008ka}, by beginning with an initial ordering and proposing and accepting new orderings $\order'$, distributed as $q(\order'|\order)$ with probability according to the Metropolis Hastings acceptance probability

\begin{equation}
p_{\textrm{acc}} = \min\left(1,\frac{p(\vect{X}|\order')p(\order')q(\order|\order')}{p(\vect{X}|\order)p(\order)q(\order'|\order)}\right),
\end{equation}

over a number of iterations. We choose to propose changes by swapping nodes in the ordering rather than more complex schemes such as ``deck cutting'', as these were found to have little impact on performance in previous studies \citep{Friedman:2003kw,Ellis:2008ka}. Proposals $\order' \sim q(\cdot|\order)$ are thus drawn by selecting two nodes within the ordering uniformly at random and exchanging their positions to produce a new ordering. In the absence of a compelling alternative, we take the prior over orderings $p(\order)$ as the uniform distribution.

Then for our emission distribution for a given state $k$ we apply a Bayesian Network ordering $\orders{k}$ generating observed data points $\vect{X}^k$ distributed as $p(\vect{X^k}|\orders{k})$ where by $\vect{X}^k$ we denote the subset of $X_{ij}$ including only rows $i$ corresponding to states $s_i=k$. 

The full method is outlined in algorithm \ref{a:gibbs}, and combines Gibbs updates of the hidden state sequence with Metropolis Hastings updates of the node orderings of the Bayesian Networks for each state at every iteration. To sample hidden state sequences and orderings from the posterior distribution the algorithm is first run for a number of burn-in iterations, after which samples are collected. Since a single iteration of our algorithm combines a full Gibbs update sweep along with an update of the Bayesian Network orderings over a number of Metropolis Hastings steps, in practice a comparatively small number of iterations of the algorithm are required to reach the posterior. To reduce the computational complexity of the Bayesian Network inference, we restrict the number of potential parents of a gene to be $2$ or less. Even in such a case, we still face a large number of possible parent sets, of size $\sum_{i=2}^{m-1}\binom{i}{2}+\sum_{i=1}^{m-1}\binom{i}{1}$, and so in the analyses presented below we restrict our data set to a subset of genes of special interest, as is commonly the case in gene  expression data analysis.

Finally, once we have inferred the hidden state sequence and generated a posterior sample of orderings corresponding to each state, we can then easily sample DAG structures from the posterior by first sampling an order from the posterior of a given state, and then sampling from the graphs consistent with this ordering, weighting the choice of parents by the local scores, and optionally attempting to account for the bias in the prior as described in \citet{Ellis:2008ka}.

\begin{algorithm}[ht]
Initialise state sequence $s_1,\ldots,s_n$\;
Set $K \leftarrow$ number of states in $s$\;
Initialise orderings $\orders{1},\ldots,\orders{K}$\;
\For{$iter \leftarrow 1$ \KwTo $N_{\textrm{iter}}$}
{
\For{$j \leftarrow 1$ \KwTo $n$}
{
\For{$k \leftarrow 1$ \KwTo $K$}
{
$l_k \leftarrow p(\orders{k}|\vect{X}^k)p(s_j=k|s_{-j},\alpha,\beta,\kappa)$\;
}
Sample new ordering $\orders{K+1}$ from $p(\order)$
$l_{K+1} \leftarrow p(\orders{K+1}|\vect{x}_j)p(s_j=K+1|s_{-j},\alpha,\beta,\kappa)$\;
$s_j \leftarrow$ sample from $1,\ldots,K+1$ with weights $\vect{l}$\;
Update $K$\;
}
Sample $\vect{\beta}|\alpha,\gamma,s_1,\dots,s_n$\;
\For{$k \leftarrow 1$ \KwTo $K$}
{
\For{$m \leftarrow 1$ \KwTo $N_{\textrm{MH}}$}
{
Sample $\orders{k}'\sim q(\orders{k}'|\orders{k})$\;
Sample $t \sim \textrm{Uniform(0,1)}$\;
$r \leftarrow \min\left(1,\frac{p(\order'|\vect{X}^k)q(\order|\order')}{p(\order|\vect{X}^k)q(\order'|\order)}\right)$\;
\If{$t<r$}{$\orders{k} \leftarrow \orders{k}'$}
}
}
}
\caption{MCMC sampler for HDP-HMM Bayesian Networks\label{a:gibbs}}
\end{algorithm}

\section{Applications}

\subsection{Example -- simulated data}

To evaluate the efficacy of our method we generated simulated data from three different Bayesian network structures and interleaved the data points into a single time series. Applying our methodology to this data we then attempted to recover the hidden state sequence.

Three different Bayesian Networks of $10$ nodes each having random structures and parameters were used, with the restriction that each node had at most two parents. Such a restriction is realistic for real world biological networks and reduces the computational complexity of the Bayesian network inference as the number of potential sets of parents of each node is greatly reduced by constraining the search. A total of $100$ data points were used, consisting of a sequence of $25$ generated by the first network, $25$ by a second network structure, another $25$ from a third network structure and finally a further $25$ data points generated by the second network structure.

The Gibbs sampling MCMC scheme outlined in algorithm \ref{a:gibbs} was applied over $500$ iterations after a $1000$ iteration burn in, with $100$ MCMC iterations of the Bayesian network order sampler run on each network structure between each Gibbs sweep.

\begin{figure}
\begin{center}
	\textbf{(a)}\includegraphics[width=3in]{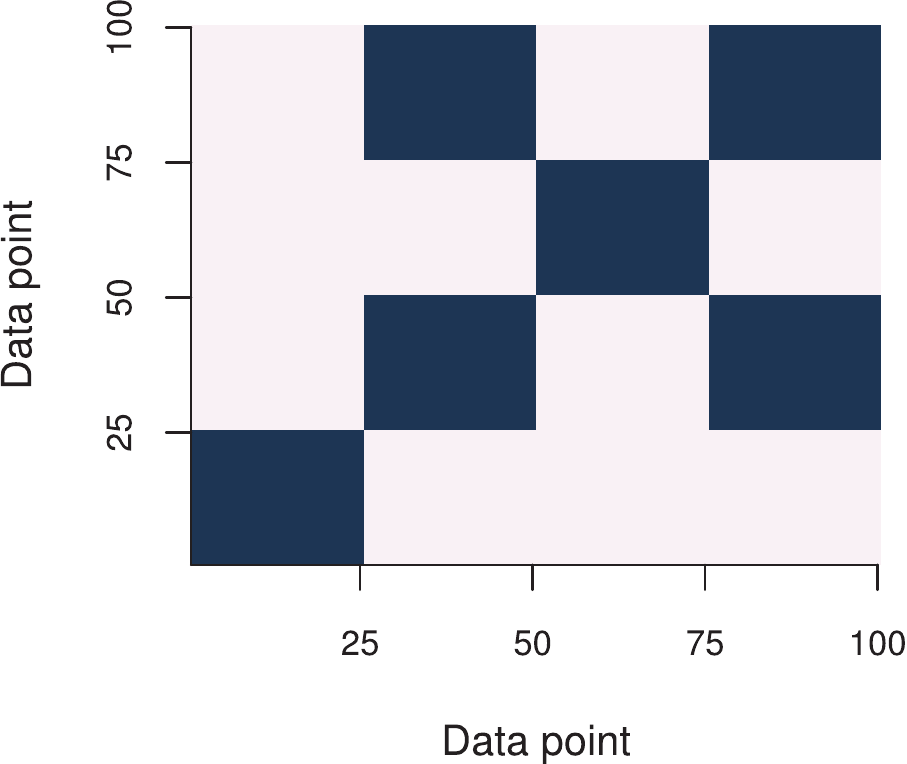}
\textbf{(b)}\includegraphics[width=3in]{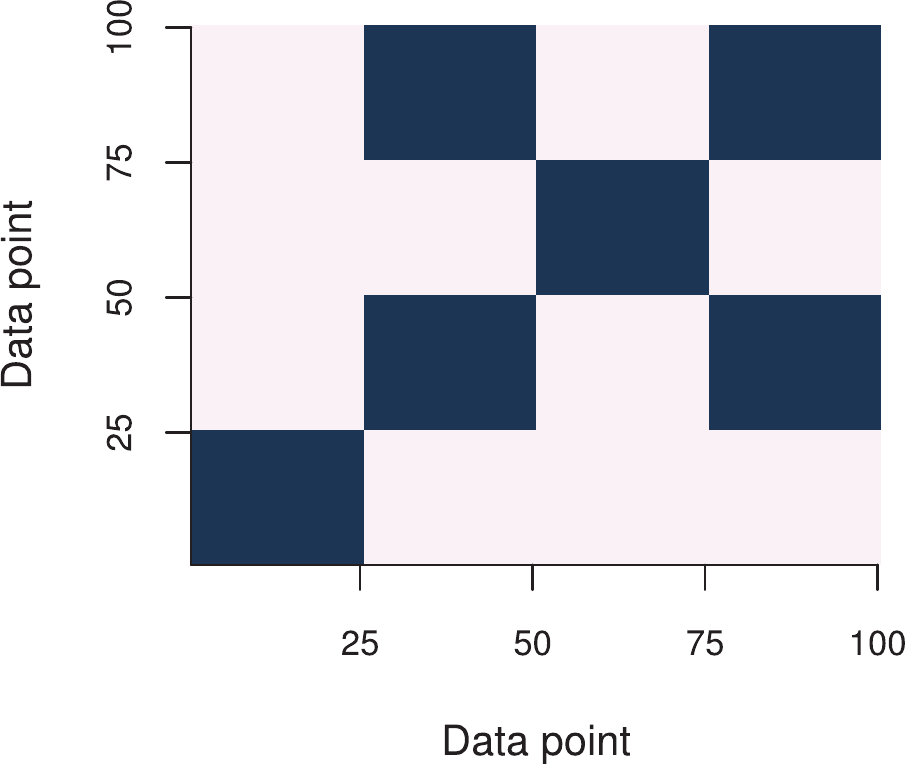}
\caption{\label{f:clust} \textbf{(a)} Counts of the number of the $500$ posterior samples in which pairs of data points where assigned to the same state. \textbf{(b)} The same illustration for the true hidden state sequence.}
\end{center}
\end{figure}

In figure \ref{f:clust} we show a comparison of the true hidden state sequence with the state sequences for the $500$ samples from the Gibbs sampler. Our method perfectly recreates the original hidden state sequence, correctly identifying that the network structure is the same between two separate segments of the sequence of data points.

\subsection{\droso midgut development gene expression data}

Applying our method to real world gene expression data, we took the publicly available gene expression data set of \citet{Li:2003ub}, as stored in the Gene Expression Omnibus database \citep{Edgar:2002tt}. This data set gives tissue specific expression levels for genes in \droso midgut at time points before and after pupariam formation, taken at $11$ time points. A subset of genes to analyse was chosen by selecting genes having the highest variance across the time series, using the genefilter R package in Bioconductor \url{www.bioconductor.org} \citep{rstats,genefilter}. This resulted in a data set of $23$ genes at $11$ time points. This allows us to apply our approach without having to consider the additional issues arising the `large-$p$-small-$n$' problem.

\begin{figure}
\begin{center}
	\includegraphics[width=3in]{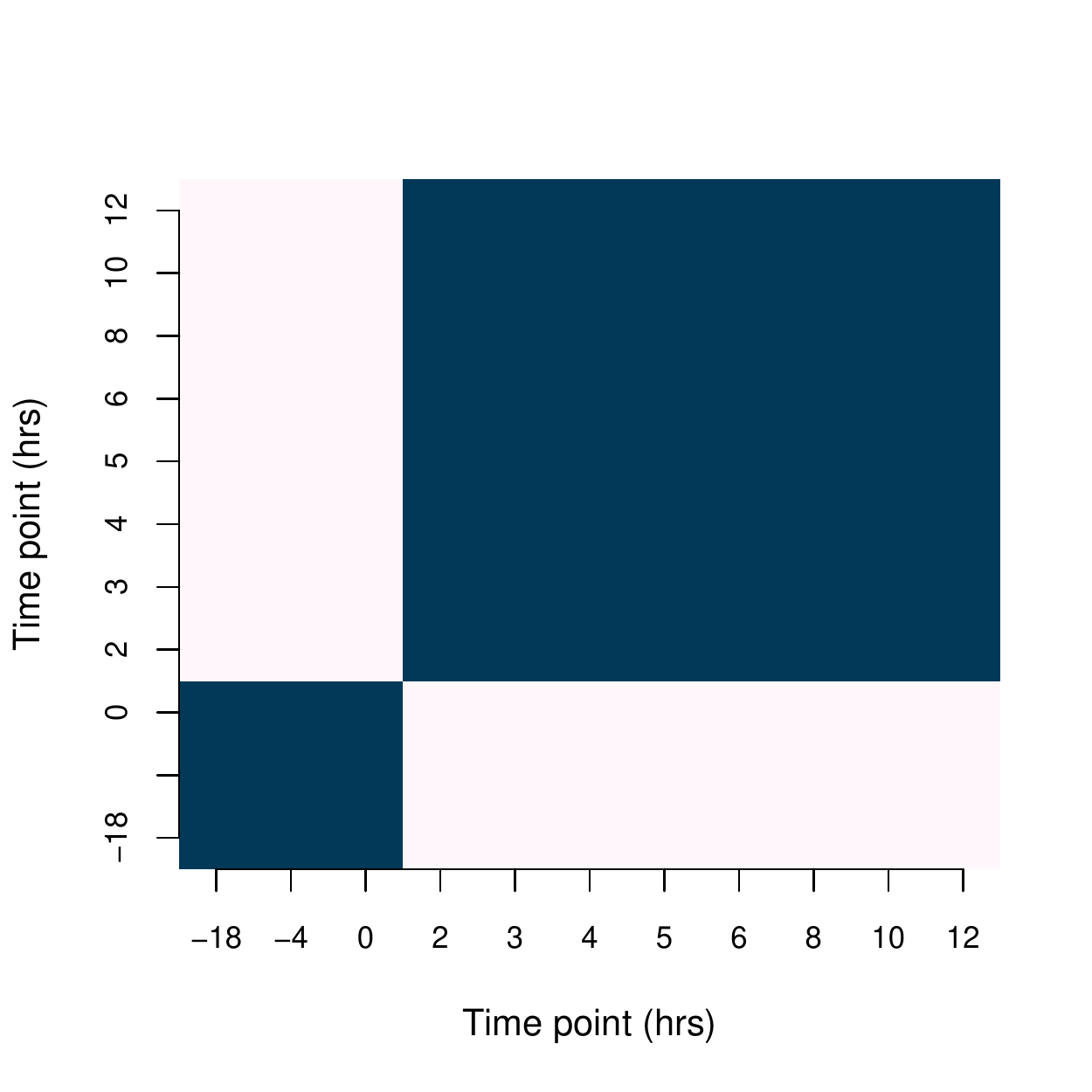}
\caption{\label{f:droso} Posterior distribution of states at each time point inferred by our method applied to the \droso midgut development expression data \citep{Li:2003ub}. States are represented by colours, and frequencies of their appearance for each time point in the posterior samples is plotted.}
\end{center}
\end{figure}

\begin{figure}
\begin{center}
\includegraphics[width=4in]{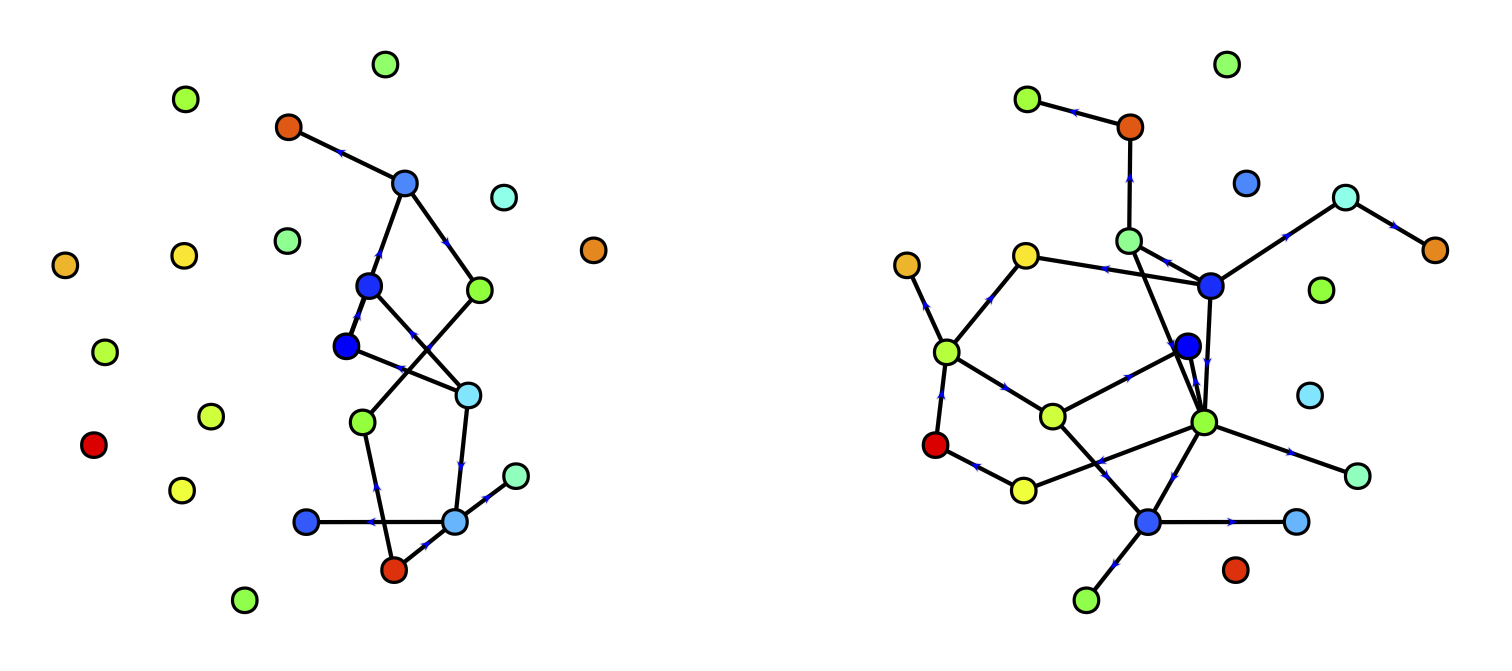}
\caption{\label{f:drosonet} Sampled Bayesian Network structures for the two states inferred by our method applied to the \droso cell cycle expression data \citep{Li:2003ub}.}
\end{center}
\end{figure}

The results shown in figure \ref{f:droso}  identify two regions of the time series having different network structures, with a change occuring after the 0 hour time point at which pupariam formation occurs. This suggests that a different structure of regulatory interactions is at work during the midgut development after the pupariam formation begins. The networks inferred for each of the different states are shown in figure \ref{f:drosonet}, illustrating a clear change between differing network structures.

\subsection{Transcriptome of starch metabolism during \araba diurnal cycle}

We have also analysed the gene expression data set of \citet{Smith:2004hu}, as included in the GeneNet \citep{Schafer:wn} R package \citep{rstats}. The data set consists of expression levels for $800$ genes encoding enzymes involved in starch synthesis and in conversion of starch to maltose and Glc, at $11$ time points over $12$ hours, transitioning from dark to light. The first 5 time points were collected during a dark period after which a switch to a light period was made, with time points spaced so that expression is measured at $0$, $1$, $2$, $4$ and $8$ hours after the switch to the dark period, and the same intervals after the switch to the light period \citep{Smith:2004hu}, as well as a final $24$ hour time point at the switch back to the dark period. A reduced subset of the $800$ genes in the data set was selected using the genefilter R package, as described previously, giving a subset of $40$ genes that were analysed using our method.

\begin{figure}
\begin{center}
\includegraphics[width=3in]{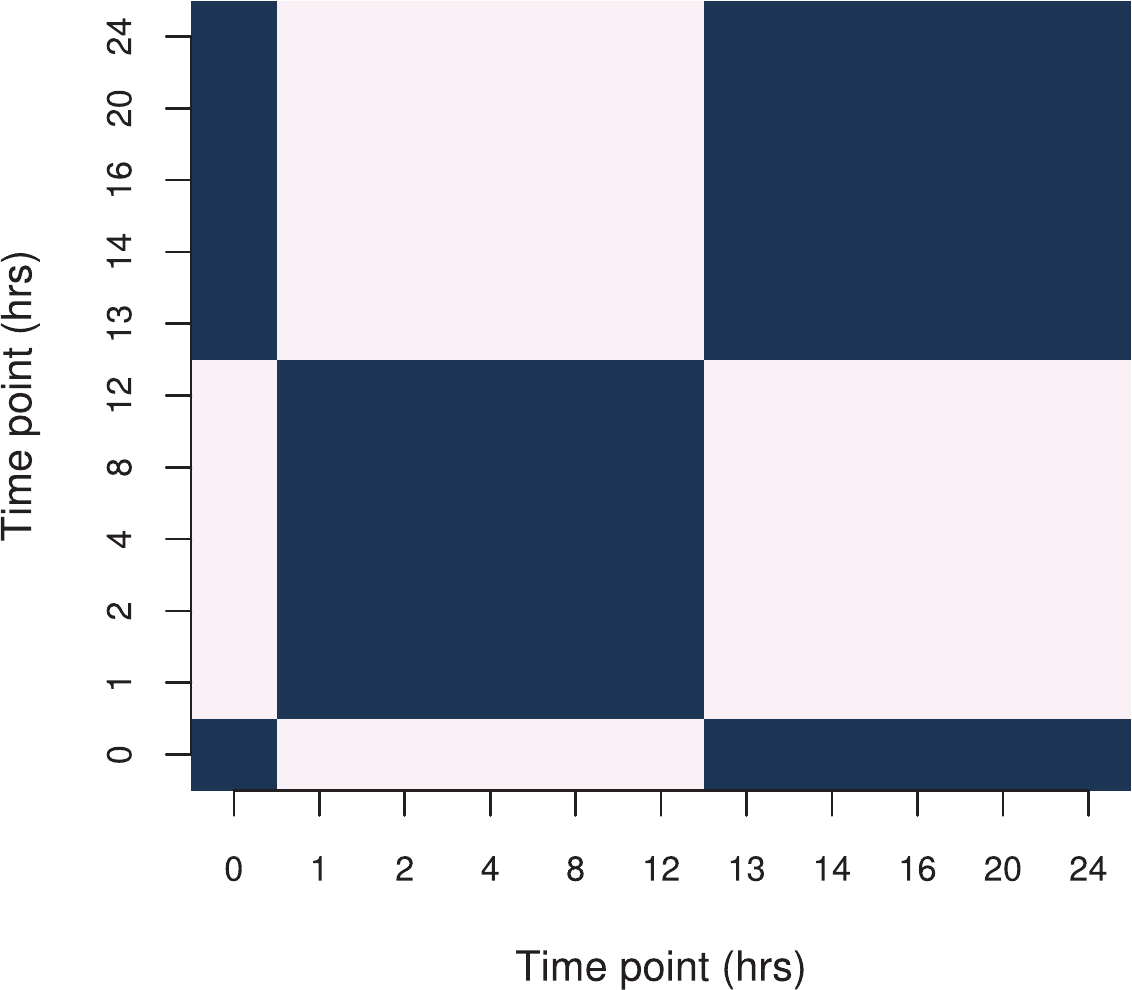}
\caption{\label{f:arth} Posterior distribution of states at each time point inferred by our method applied to the \araba diurnal cycle expression data \citep{Smith:2004hu}. States are represented by colours, and frequencies of their appearance for each time point in the posterior samples is plotted.}
\end{center}
\end{figure}

\begin{figure}
\begin{center}
\includegraphics[width=4in]{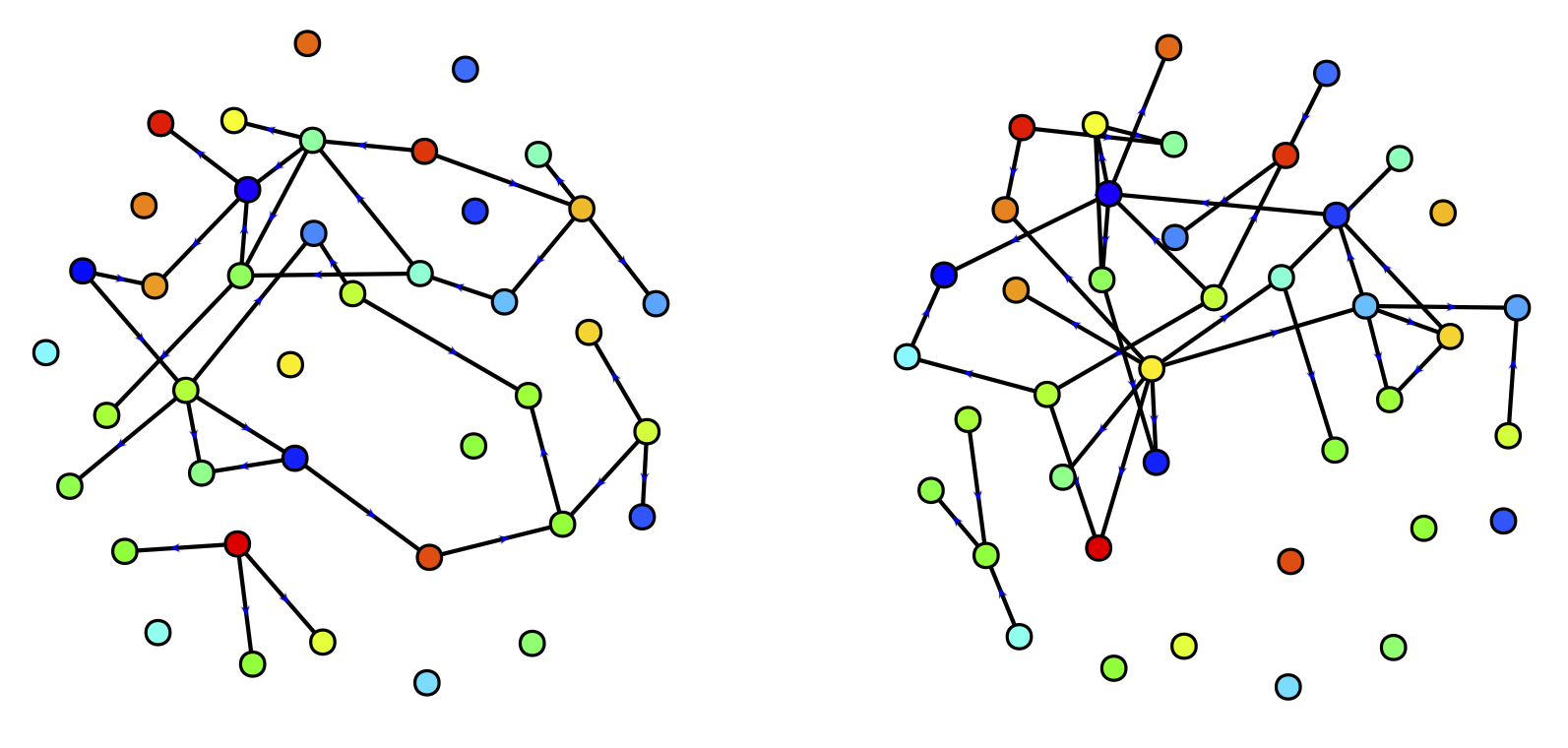}
\caption{\label{f:arthnet} Sampled Bayesian Network structures for the two states inferred by our method applied to the \araba diurnal cycle expression data \citep{Smith:2004hu}.}
\end{center}
\end{figure}

In figure \ref{f:arth} we show the results generated by our method, clearly indicating two distinct phases within the time series. It appears that one phase is detected from $1$ to $12$ hours, with a second phase inferred between $13$ and $24$ hours, that is also represented at the initial time point. This is consistent with the design of the experiment, as a change in expression would perhaps not be expected to be observed immediately at the point at which the switch between light and dark takes place, but rather later at a subsequent time point, as is observed here. Since the $24$ hour time point was taken under the same conditions as the initial time point, one would expect these two time points to be grouped together using our method. The networks inferred for the two different phases, shown in figure \ref{f:arthnet}, again demonstrate a clear change in the network structure, with the two networks having distinct topologies.

\section{Discussion}

From our simulated data it is clear that the HDP-HMM Bayesian Network sampler we have constructed accurately infers the hidden state sequences governing Bayesian Networks that capture how the regulatory organisation of a biological system changes with time. The accuracy of our method on test data suggests that it will perform well on real world data sets, whilst the existence of more sophisticated and demonstrably more efficient samplers indicates that there is room for even further improvement and computational efficiency. For example the beam sampler of \citet{VanGael:2008fs} and the Hierarchical Chinese Restaurant Process (HCRP) formalism of \citet{Makino:2011wy} have better mixing and perform better than standard Gibbs samplers, especially on time series such as those we examine here where neighbouring states are likely to be correlated.  

We remark that it is essential to consider the fluid nature of regulatory network structures when inferring networks from data sets where such change is likely. Performing an analysis on data using a model with a fixed network structure, when it is known that the network structure will change due to some stimulus, is inherently incorrect, and thus will introduce unnecessary bias into the results. Whilst it may be possible to infer correct results from an incorrect model, it would not seem wise to rely on such approaches when alternatives exist.

Our methodology crucially accounts for the sequential nature of the data, something that has previously been overlooked \citep{Grzegorczyk:2008ie,Ickstadt2011}, but we feel is crucial to the modelling of gene expression time series data sets. Furthermore our methodology has an advantage over changepoint models that data may be shared between distinct segments of the time series sharing the same hidden state when inferring the network structure -- something that is explicitly represented in our model, but generally omitted in changepoint models. This is especially import in gene expression data analysis where time points are a scarce and valuable resource.

The versatility of the HDP-HMM means that our methodology is applicable not only to time series data where the underlying process is divided into distinct contiguous segments, as would be expected in gene regulatory networks, but also to processes describable by a Markov process, for example rapidly changing between a sequence of hidden states with some underlying transition mechanism. Thus it may be of use for other problems of network inference in systems biology outside of the area of sequential gene expression time series data, or in other fields where networks that change with time are encountered.

Finally whilst other methods may require manual specification of an appropriate prior distribution on the number of possible states, taking a nonparametric approach allows our prior distribution to naturally expand to explain the observed data as the size and complexity of the data grows. Bayesian nonparametric methods demonstrably outperform regular priors in a variety of applications and we have shown here their potential in modelling hidden variables in theoretical systems biology. 

\bibliographystyle{nat}
\bibliography{refs}

\begin{thebibliography}{27}
\providecommand{\natexlab}[1]{#1}

\bibitem[{Beal \emph{et~al.}(2002)Beal, Ghahramani and Rasmussen}]{Beal:2002tt}
Beal, M., Ghahramani, Z. and Rasmussen, C.E.
\newblock {The infinite hidden Markov model}.
\newblock \emph{Advances in neural information Processing Systems}, 2002.

\bibitem[{Edgar \emph{et~al.}(2002)Edgar, Domrachev and Lash}]{Edgar:2002tt}
Edgar, R., Domrachev, M. and Lash, A.E.
\newblock {Gene Expression Omnibus: NCBI gene expression and hybridization
  array data repository.}
\newblock \emph{Nucleic acids research}, 30(1):207--210, 2002.

\bibitem[{Ellis and Wong(2008)}]{Ellis:2008ka}
Ellis, B. and Wong, W.H.
\newblock {Learning causal Bayesian network structures from experimental data}.
\newblock \emph{Journal of the American Statistical Association},
  103(482):778--789, 2008.

\bibitem[{Fox \emph{et~al.}(2008)Fox, Sudderth, Jordan and
  Willsky}]{Fox:2008dm}
Fox, E.B., Sudderth, E.B., Jordan, M.I. and Willsky, A.S.
\newblock {An HDP-HMM for systems with state persistence}.
\newblock In \emph{ICML '08: Proceedings of the 25th international conference
  on Machine learning}. ACM, 2008.

\bibitem[{Fox \emph{et~al.}(2009)Fox, Sudderth, Jordan and
  Willsky}]{Fox:2009kf}
Fox, E.B., Sudderth, E.B., Jordan, M.I. and Willsky, A.S.
\newblock {A sticky HDP-HMM with application to speaker diarization}.
\newblock \emph{arXiv.org}, stat.ME, 2009.

\bibitem[{Friedman and Koller(2003)}]{Friedman:2003kw}
Friedman, N. and Koller, D.
\newblock {SpringerLink - Machine Learning, Volume 50, Numbers 1-2}.
\newblock \emph{Machine Learning}, 50(1/2):95--125, 2003.

\bibitem[{Geiger and Heckerman(2002)}]{Geiger:2002uj}
Geiger, D. and Heckerman, D.
\newblock {Parameter Priors for Directed Acyclic Graphical Models and the
  Characterization of Several Probability Distributions}.
\newblock \emph{Annals of statistics}, 2002.

\bibitem[{Gentleman \emph{et~al.}(2007)Gentleman, Carey, Huber and
  Hahne}]{genefilter}
Gentleman, R., Carey, V., Huber, W. and Hahne, F.
\newblock \emph{genefilter: genefilter: methods for filtering genes from
  microarray experiments}, 2007.
\newblock R package version 1.34.0.

\bibitem[{Grzegorczyk and Husmeier(2008)}]{Grzegorczyk:2008tw}
Grzegorczyk, M. and Husmeier, D.
\newblock {Improving the structure MCMC sampler for Bayesian networks by
  introducing a new edge reversal move}.
\newblock \emph{Machine Learning}, 71(2-3), 2008.

\bibitem[{Grzegorczyk \emph{et~al.}(2008)Grzegorczyk, Husmeier, Edwards, Ghazal
  and Millar}]{Grzegorczyk:2008ie}
Grzegorczyk, M., Husmeier, D., Edwards, K.D., Ghazal, P. and Millar, A.J.
\newblock {Modelling non-stationary gene regulatory processes with a
  non-homogeneous Bayesian network and the allocation sampler.}
\newblock \emph{Bioinformatics (Oxford, England)}, 24(18):2071--2078, 2008.

\bibitem[{Ickstadt(2011)}]{Ickstadt2011}
Ickstadt, K.
\newblock {Nonparametric Bayesian Networks (with discussion)}.
\newblock In J.~Bernardo, M.J. Bayarri, J.O. Berger, A.P. Dawid, D.~Heckerman,
  A.F.M. Smith and M.~West, editors, \emph{Bayesian Statistics 9}, pages
  135--155. Oxford University Press, 2011.

\bibitem[{Koski and Noble(2009)}]{Koski:1315082}
Koski, T. and Noble, J.
\newblock \emph{{Bayesian Networks: An Introduction (Wiley Series in
  Probability and Statistics)}}.
\newblock Wiley, 1st edition, 2009.

\bibitem[{L{\`e}bre(2009)}]{Lebre:2009ea}
L{\`e}bre, S.
\newblock {Inferring dynamic genetic networks with low order independencies.}
\newblock \emph{Statistical applications in genetics and molecular biology},
  8(1):Article 9--, 2009.

\bibitem[{L{\`e}bre \emph{et~al.}(2010)L{\`e}bre, Becq, Devaux, Stumpf and
  Lelandais}]{Lebre:2010jz}
L{\`e}bre, S., Becq, J., Devaux, F., Stumpf, M.P.H. and Lelandais, G.
\newblock {Statistical inference of the time-varying structure of
  gene-regulation networks.}
\newblock \emph{BMC systems biology}, 4:130--, 2010.

\bibitem[{Li and White(2003)}]{Li:2003ub}
Li, T.R. and White, K.P.
\newblock {Tissue-specific gene expression and ecdysone-regulated genomic
  networks in Drosophila.}
\newblock \emph{Developmental cell}, 5(1):59--72, 2003.

\bibitem[{Madigan \emph{et~al.}(1995)Madigan, York and Allard}]{Madigan:1995vv}
Madigan, D., York, J. and Allard, D.
\newblock {Bayesian Graphical Models for Discrete Data}.
\newblock \emph{International Statistical Review}, 63(2):215--232, 1995.

\bibitem[{Makino \emph{et~al.}(2011)Makino, Takei, Sato and
  Mochihashi}]{Makino:2011wy}
Makino, T., Takei, S., Sato, I. and Mochihashi, D.
\newblock {Restricted Collapsed Draw: Accurate Sampling for Hierarchical
  Chinese Restaurant Process Hidden Markov Models}.
\newblock \emph{arXiv.org}, stat.ML, 2011.

\bibitem[{Opgen-Rhein and Strimmer(2007)}]{OpgenRhein:2007ih}
Opgen-Rhein, R. and Strimmer, K.
\newblock {From correlation to causation networks: a simple approximate
  learning algorithm and its application to high-dimensional plant gene
  expression data.}
\newblock \emph{BMC systems biology}, 1:37, 2007.

\bibitem[{{R Development Core Team}(2011)}]{rstats}
{R Development Core Team}.
\newblock \emph{R: A Language and Environment for Statistical Computing}.
\newblock R Foundation for Statistical Computing, Vienna, Austria, 2011.
\newblock {ISBN} 3-900051-07-0.

\bibitem[{Rodriguez \emph{et~al.}(2010)Rodriguez, Lenkoski and
  Dobra}]{2010arXiv1001.4208R}
Rodriguez, A., Lenkoski, A. and Dobra, A.
\newblock {Sparse covariance estimation in heterogeneous samples}.
\newblock \emph{arXiv.org}, stat.ME:4208, 2010.

\bibitem[{Schafer and Opgen-Rhein(2006)}]{Schafer:wn}
Schafer, J. and Opgen-Rhein, R.
\newblock {Reverse engineering genetic networks using the GeneNet package},
  2006.

\bibitem[{Sch{\"a}fer and Strimmer(2005)}]{Schafer:2005kl}
Sch{\"a}fer, J. and Strimmer, K.
\newblock {An empirical Bayes approach to inferring large-scale gene
  association networks.}
\newblock \emph{Bioinformatics (Oxford, England)}, 21(6):754--764, 2005.

\bibitem[{Sethuraman(1994)}]{Sethuraman:1994}
Sethuraman, J.
\newblock {A constructive definition of Dirichlet priors}.
\newblock \emph{Statistica Sinica}, 4:639--650, 1994.

\bibitem[{Smith \emph{et~al.}(2004)Smith, Fulton, Chia, Thorneycroft, Chapple,
  Dunstan, Hylton, Zeeman \emph{et~al.}}]{Smith:2004hu}
Smith, S., Fulton, D., Chia, T., Thorneycroft, D., Chapple, A., Dunstan, H.,
  Hylton, C., Zeeman, S. \emph{et~al.}
\newblock {Diurnal changes in the transcriptome encoding enzymes of starch
  metabolism provide evidence for both transcriptional and posttranscriptional
  regulation of starch metabolism in arabidopsis leaves}.
\newblock \emph{Plant Physiology}, 136(1):2687--2699, 2004.

\bibitem[{Teh and Jordan(2010)}]{Teh:2010to}
Teh, Y.W. and Jordan, M.I.
\newblock {Hierarchical Bayesian nonparametric models with applications}.
\newblock In \emph{Bayesian nonparametrics}, pages 158--207. Cambridge Univ.
  Press, Cambridge, 2010.

\bibitem[{Teh \emph{et~al.}(2006)Teh, Jordan and Beal}]{Teh:2006wl}
Teh, Y., Jordan, M. and Beal, M.
\newblock {Hierarchical dirichlet processes}.
\newblock \emph{Journal of the American Statistical Association}, 2006.

\bibitem[{Van~Gael \emph{et~al.}(2008)Van~Gael, Saatci, Teh and
  Ghahramani}]{VanGael:2008fs}
Van~Gael, J., Saatci, Y., Teh, Y.W. and Ghahramani, Z.
\newblock {Beam sampling for the infinite hidden Markov model}.
\newblock In \emph{ICML '08: Proceedings of the 25th international conference
  on Machine learning}. ACM, 2008.

\end{thebibliography}
\end{document}